# Integrated Information as a Metric for Group Interaction: Analyzing Human and Computer Groups Using a Technique Developed to Measure Consciousness

David Engel[1,2,3] and Thomas W. Malone[1,2,*]

[1] Massachusetts Institute of Technology, Center for Collective Intelligence, Cambridge, MA 02142, USA

[2] Massachusetts Institute of Technology, Sloan School of Management, Cambridge, MA 02142, USA

[3] Now at Google, Inc., Zurich, Switzerland

[*] Corresponding author.  Email:  malone@mit.edu

**Abstract**.  Researchers in many disciplines have previously used a variety of mathematical techniques for analyzing group interactions.  Here we use a new metric for this purpose, called "integrated information" or "phi."  Phi was originally developed by neuroscientists as a measure of consciousness in brains, but it captures, in a single mathematical quantity, two properties that are important in many other kinds of groups as well:  differentiated information and integration.  Here we apply this metric to the activity of three types of groups that involve people and computers.  First, we find that 4-person work groups with higher measured phi perform a wide range of tasks more effectively, as measured by their collective intelligence.  Next, we find that groups of Wikipedia editors with higher measured phi create higher quality articles. Last, we find that the measured phi of the collection of people and computers communicating on the Internet increased over a recent six-year period.  Together, these results suggest that integrated information can be a useful way of characterizing a certain kind of interactional complexity that, at least sometimes, predicts group performance.  In this sense, phi can be viewed as a potential metric of effective group collaboration.  Since the metric was originally developed as a measure of consciousness, the results also raise intriguing questions about the conditions under which it might be useful to regard groups as having a kind of consciousness.

## Introduction

A vast number of phenomena in the world arise out of the interactions of individuals in groups, from the emotional tone of a family [1,2] to the productivity of an economy [3] to the spread of disease in a community [4], and researchers in a variety of disciplines have used many different mathematical tools to analyze these phenomena.  For instance, psychologists have used Markov models to analyze the sequences of actions in small groups of people [5–7], economists have used general equilibrium theory to analyze the interactions among buyers and sellers in a market [8], and sociologists have used graph theory to analyze various kinds of social networks [4,9].

In this paper, we examine another mathematical technique that has not previously been used for analyzing group interactions. This technique, based on information theory, is intriguing because it was developed as a physical measure that would correlate with the consciousness of a brain [10–14]. We will see, however, that the metric is general enough to apply to many other kinds of systems, and we focus here on using it to analyze groups of people and computers.

*What is integrated information?*

The metric we use is called "integrated information" or "phi" and was proposed by Tononi and colleagues [10–14]. There have been several successively refined versions of phi (summarized in [12]), but all the versions aim to quantify the integrated information in a system. Loosely speaking, this means the amount of information generated by the system as a whole that is more than just the sum of its parts. The phi metric does this by splitting the system into subsystems and then calculating how much information can be explained by looking at the system as a whole but not by looking at the subsystems separately.

In other words, for a system to have a high value of phi, it must, first of all, generate a large amount of *information*. Information can be defined as the reduction of uncertainty produced when one event occurs out of many possible events that might have occurred [15]. Thus a system can produce more information when it can produce more possible events. This, in turn, is possible when it has more different parts that can be in more different combinations of states. In other words, a system needs a certain kind of differentiated complexity in its structure in order to generate a large amount of information.

But phi requires more than just information; it also requires the information to be *integrated* at the level of the system as a whole. A system with many different parts could produce a great deal of information, but if the different parts were completely independent of each other, then the information would not be integrated at all, and the value of phi would be 0. For a system to be integrated, the events in some parts of the system need to depend on events in other parts of the system. And the stronger and more widespread these interdependencies are, the greater the degree of integration.

For instance, a single photodiode that senses whether a scene is light or dark does not generate much information because it can only be in two possible states. But even a digital camera with a million photodiodes, which can discriminate among $2^{1,000,000}$ possible states, would not produce any integrated information because each photodiode is independently responding to a different tiny segment of the scene. Since there are no interdependencies among the different photodiodes, there is no integrated information [13].

Tononi and colleagues argue that these two properties—differentiated information and integration—are both essential to the subjective experience of consciousness. For example, the conscious perception of a red triangle is an integrated subjective experience that is more than the sum of perceiving "a triangle but no red, plus a red patch but no triangle" [12]. The information is integrated in the sense that we cannot consciously perceive the triangle's shape independently from its color, nor can we perceive the left visual hemisphere independently from the right. Said differently, integrated information in conscious experience results from functionally specialized subsystems that interact significantly with each other [16].

Even though there is not yet a general consensus among researchers that phi actually measures consciousness in humans (e.g., [17]), it does capture these important aspects of the experience of consciousness, and the mathematical behavior of phi is also consistent with many empirical observations of human consciousness [10,18–22].

Interestingly, the two properties used to define phi—differentiated information and integration—are similar to properties that are important in many other kinds of systems, too. For example, Adam Smith [23] observed that economic systems are often more productive when (a) division of labor leads different people to specialize in different kinds of work and (b) the "invisible hand" of the market integrates their diverse efforts. Lawrence and Lorsch [24] discussed the importance of differentiation and integration in large, hierarchical human organizations: (a) dividing the organization into specialized subunits and (b) integrating these units to achieve the goals of the overall organization. And in many fields of engineering and other kinds of design, effective problem solving often involves (a) dividing a problem into subparts and (b) integrating solutions for the subparts into a solution for the whole problem [25–27].

In other words, the mathematical concept of integrated information provides a quantitative way of measuring a combination of two properties that are important across a wide range of different types of systems. And whether phi is measuring consciousness or not, it is clearly measuring something that is of potential interest to many different disciplines.

*A mathematical formulation of integrated information*

The concept of integrated information, or phi, can be represented mathematically as follows [28]:

$$\emptyset = \sum_{k=1}^{r} H(M_0^k \mid M_1^k) - H(X_0 \mid X_1), \quad (1)$$

where H(X|Y) is the entropy of variable X given knowledge of variable Y, $X_0$ and $X_1$ are the states of the whole system at time $t_0$ and $t_1$, respectively, and $M_0^k$ and $M_1^k$ are subsets of X that completely partition the parts of X at these times. For example, $H(M_0^k \mid M_1^k)$ quantifies how much of the uncertainty of subsystem k at time $t_0$ cannot be explained by knowledge of the state of the subsystem at time $t_1$.

Summing over all subsystems (the first term in equation (1)) gives us the amount of entropy that cannot be explained by the subsystems themselves. The second term in equation (1) quantifies the conditional entropy of the whole system. Thus phi is high if there is a large amount of entropy that cannot be explained by looking at the subsystems separately but that is explained by looking at the system as a whole.

The value calculated by equation (1) is the phi as defined by Tononi and colleagues if and only if the partitioning is chosen as the maximum information bipartition (MIB), that is, the decomposition into two parts that are most independent. More thorough descriptions of phi can be found in [11–14,21,29].

*Applying the phi metric*

As we saw above, phi provides a quantitative measure of properties that figure prominently in several theories of group performance, so we first test whether it is, in fact, correlated with performance in two different kinds of groups: (a) small groups of experimental subjects working together on shared laboratory tasks, and (b) groups of Wikipedia contributors improving Wikipedia articles over time. As a further test of the applicability of phi, we also examine whether it detects what we might assume would be the increasingly differentiated and integrated complexity of the Internet over time. We evaluate this by applying the phi metric to data about all the computers (and people) communicating over a specific Internet backbone during a six-year period.

In order to apply the theoretical definition of phi, a complete model of the rules governing state transitions in the system is needed, but such models are rarely available for observational data. Therefore, we use two alternative versions of phi suggested by Barrett and Seth [28] that estimate conditional probabilities for state transitions from the actual observed data (see Methods). In each case, in order to use the phi metric, we needed to determine: (a) a characterization of the state of the system at different times, and (b) a time delay with respect to which phi will be calculated [10].

**Results**

*Study 1: Small work groups*

In Study 1, we applied the phi metric to interaction data we gathered from a previous study of groups performing a series of tasks designed to measure their collective intelligence [30]. Collective intelligence (CI) is a statistical factor for a group that predicts the group's performance on a wide range of tasks, just as individual intelligence does for individuals [31]. Following [31], we measured collective intelligence by asking 68 groups of 4 people each to perform a range of tasks, from brainstorming and memory tasks to solving visual and word puzzles. Then we did a factor analysis of the groups' performance on these tasks. Analogously to individual intelligence, we called the first factor that emerged from this factor analysis a group's collective intelligence.

All groups did the tasks using a shared online tool. In the first condition, the groups also communicated face-to-face while collaborating on the tasks. In the second condition, groups only communicated via online text chat.

In order to apply the phi metric, we characterized the state of the group in terms of which group member was communicating at which point in time. This yielded a binary state vector for each person (talking or not talking) in each time step (see Methods). In the face-to-face condition, we recorded separate audio files for each person and encoded which individuals were talking at each time. For the online groups we analyzed the chat log to determine which group members sent messages at each time.

We also needed to determine the time delay with respect to which phi would be calculated. For the online condition, we expect to see an influence of what is said in one comment on the next comment, so we set the time delay to one "timestep," that is, the time from one textual comment to the next one.

For the face-to-face groups, we don't expect the actions of one group member to immediately influence the actions of another one. Instead, we would expect time delays on the order of a few seconds, the approximate time it takes for a person to hear and respond to what someone else says. To determine exactly how long this delay should be, we plotted average phi for different time delays (Fig 1). There is a clear peak at around 2 seconds, an intuitively plausible value, so we used this value as the time delay for analyzing phi.

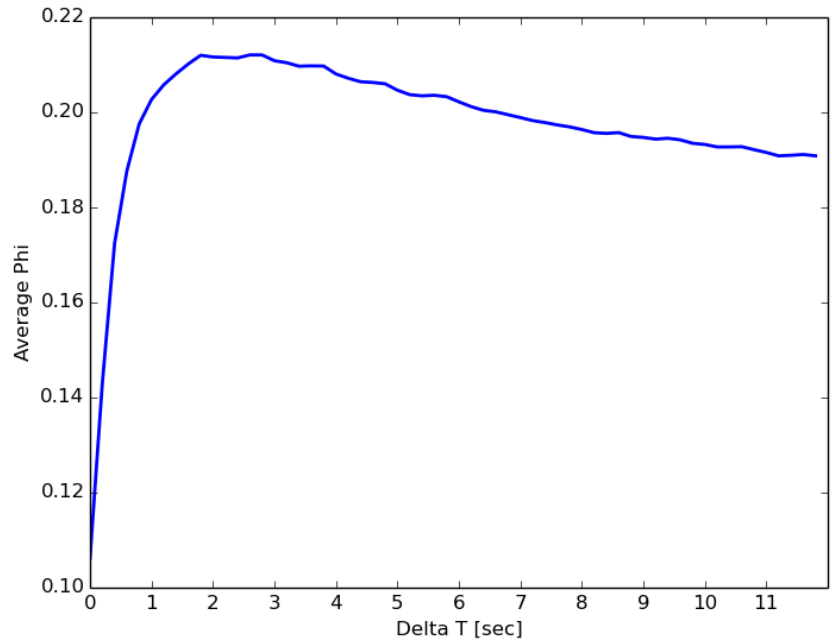

**Fig 1**. Average phi for face-to-face groups computed with different time delays.

When we calculated phi, it was significantly correlated with the measured collective intelligence of the groups in both the face-to-face condition (r=0.401, p=0.047) and the online condition (r=0.352, p=0.035). The correlation was also very significant when we normalized both phi and CI scores and combined both conditions (r=0.372, p=0.003).

*Study 2: Groups of Wikipedia editors*

In Study 2, we analyzed the edit history of the articles from the "vital articles" list in Wikipedia [32] which at the time of retrieval contained 1000 articles. The quality of Wikipedia articles are classified by the community into the classes FA (Featured Article), which is the highest class, followed, in order of decreasing quality, by the categories A, GA (Good Article), B and C [33].

We examined all edits made in the 30-, 60-, and 90-day periods before an article was promoted to its current quality level, discarding the periods in which only one or two editors were active (see descriptive statistics in S3 Table). All editors who edited an article during a given time period were considered members of the "group" for that article.

Since the number of editors in these groups can be large (up to several hundred), computing phi for the minimum information bipartition is no longer computationally feasible since it would entail testing all possible bipartitions. Consequently we used an alternative measure of phi that treats each node as a separate partition (see Methods).

Using this measure, it is clear that, in general, groups of editors who produce higher quality articles also have significantly higher phi (Fig 2 and S1 Fig; 30 days: Kendall Tau 0.0697, $p < 0.00001$; 60 days: Kendall Tau 0.113; $p < 0.0001$; 90 days Kendall Tau 0.1267, $p < 0.00001$, respectively).

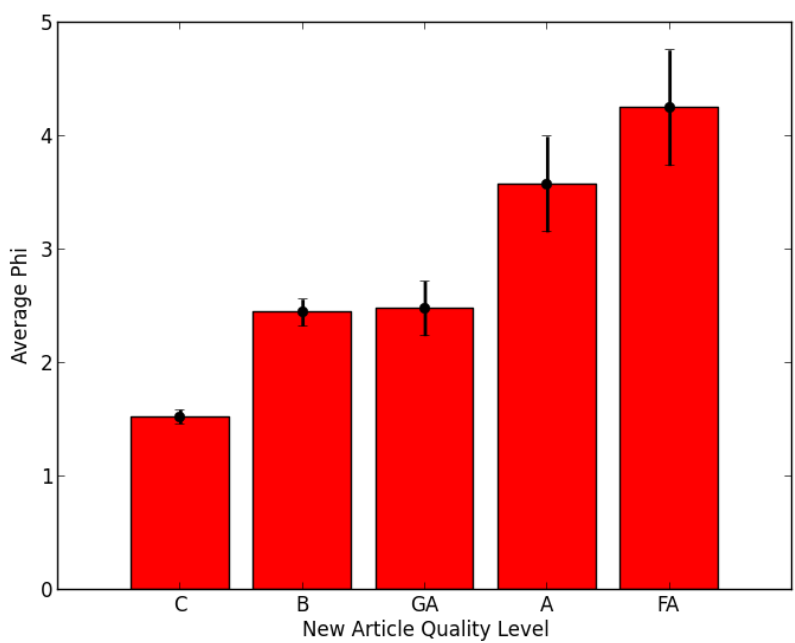

**Fig. 2**: Average phi for groups editing Wikipedia articles of different quality levels in the 60-day period before the articles were promoted to their current quality level. Quality levels are arranged in order of increasing quality. Error bars show standard error.

To test whether this effect on phi is a result of the number of editors or of the number of edits they made, we created a regression model that predicted phi for each article from the number of editors, the number of edits per editor and four variables encoding the quality of the article. Note, that we used the categorical variables with standard treatment contrasts. That means the first level (quality C) was used as a reference and did not get parameter estimation. All other categorical quality parameters were computed relative to quality C. We did not expect the quality levels that are not significantly different from each other (see above) to be separate significant factors by themselves, and we indeed found that only quality levels B and A are significant predictors of phi (Table 1).

|  | **Estimate** | **Std. Error** | **t Value** | **Pr(>\|t\|)** |
|---|---|---|---|---|
| (Intercept) | 1.594 | 0.207 | 7.708 | 1.75e-14 *** |
| Number of editors | 0.012 | 0.001 | 18.076 | < 2e-16 *** |
| Edits per editor | -0.123 | 0.009 | -14.46 | < 2e-16 *** |
| Quality B | 0.604 | 0.240 | 2.515 | 0.012 * |
| Quality GA | 0.345 | 0.311 | 1.11 | 0.267 |
| Quality A | 1.291 | 0.359 | 3.592 | 0.0003 *** |
| Quality FA | 0.370 | 0.378 | 0.978 | 0.328 |

Adjusted R-squared = 0.127, F = 70.11, p < 2.2e-16

**Table 1:** Regression results when predicting phi for each article from number of edits, average number of edits per editor, and newly acquired quality level of the article.

The effect of the newly attained quality level in the presence of covariates was also assessed by a likelihood ratio test between the model without quality as a variable and the full model. This showed that the article quality was still significantly correlated with phi, even when controlling for the other factors (F = 3.6847, p = 0.0053).

More specifically, pairwise Wilcoxon ranksum tests show that the groups editing FA and A articles have significantly higher phi values than GA and B articles which, in turn, are significantly higher than C (Wilcoxon z-Statistics = 5.6024 p < 0.00001 between C and B and z-statistic = 3.5132, p = 0.0004 between GA and A).

*Study 3: Groups of computers and people on the Internet*

In Study 3, we applied the phi metric to a sample of the Internet traffic that passed through one Internet backbone over a six-year period [34]. We sampled one-minute segments of traffic separated by approximately six-month intervals over this period (see Methods). We encoded the state of the system in terms of whether a given machine was active (i.e., sent a data packet) at a given time. We picked a time delay of 1 time step and chose the time step size that maximized phi averaged over all years in the dataset (see Fig 3, Methods). In this case, the maximum is at 100 ms which is reassuring since it coincides well with average response times observed over the Internet (see Methods). In order to keep the computations of phi tractable, we also reduced the number of machines by randomly sampling subsets of machines that communicated with each other (see Methods).

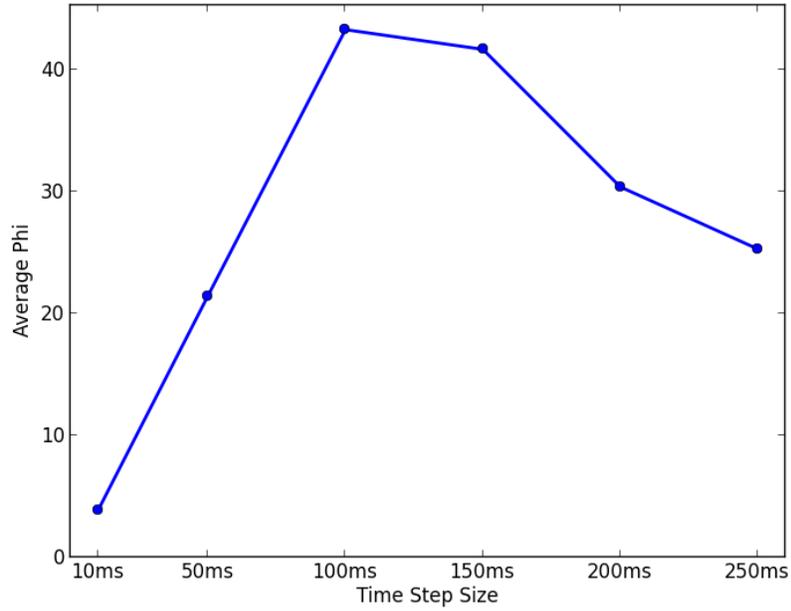

**Fig. 3.** Average phi for Internet traffic computed with different time step sizes.

When computing phi, there appears to be a steady upward trend over time. For example, Fig 4 shows one example of a highly significant relationship between the date and phi ($\beta = 1.779$, $p<0.0001$). Similar results were obtained for numerous other sampling methods and parameters (see Methods, S2 – S4 Figs). These results all take into account adjustments for a discontinuity between 2011 and 2012 due to a change in the hardware configuration at the recording site [35]. It is important to note that the results do not arise simply from an increasing number of machines in the Internet over time, since the number of machines in the samples analyzed is constant in each case (see Methods).

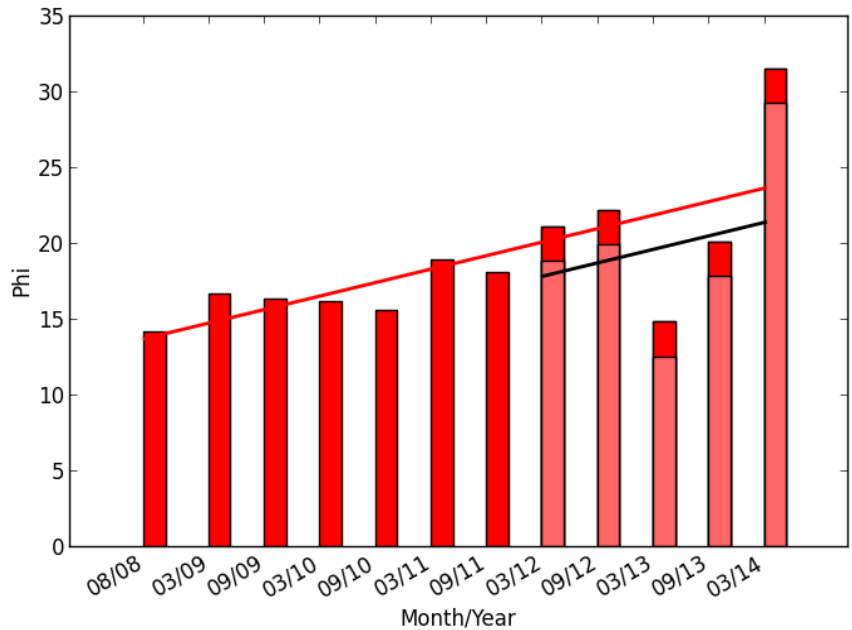

**Fig. 4**: Average phi computed on Internet traffic data over a span of 6 years (node sampling = random walk, node sample size = 100, time step size = 100 ms. See details in Methods). A change in hardware at the recording site between 2011 and 2012 caused a drop in subsequent recorded traffic **[35]**. The actual traffic in subsequent years is indicated by a horizontal black line and light red bars. The red bars and the red line show values adjusted to compensate for this change (see Methods).

**Discussion**

Together, these results suggest that the concept of integrated information, as formalized by the phi metric, can be usefully applied to group interactions. To begin with, the time delays at which this measure is maximized are intuitively plausible for a measure of interaction: 2 seconds for face-to-face human groups and 100 ms for machines on the Internet.

*Predicting group performance*

More importantly, phi is correlated with various measures of group performance. In 4-person work groups, it is correlated with the groups' collective intelligence. Previous work has shown that collective intelligence, in turn, predicts a group's performance on a wide range of other tasks [30,31,36]. Furthermore, in groups of Wikipedia editors, phi is correlated with the quality of the articles the groups edited.

Since phi can be calculated from a relatively small sample of group interactions, this suggests that it might be possible to predict many kinds of group performance, long before a group's output is complete, merely by measuring phi. This possible use of phi seems plausible because we could interpret phi as a measure of *group collaboration*, and it seems likely that the degree of collaboration in a group could be a good predictor of the group's performance in many situations.

This use of phi would be analogous to the use of intelligence tests for individuals [37] or groups [31] to predict performance on future tasks. But these intelligence tests are *interventional* measures; they require people to do specific testing activities they would not otherwise have done in order to predict their performance on another task. The versions of phi used here, on the other hand, are *observational* measures; as we have seen, they can be calculated merely by observing what people are doing anyway. In this sense, then, phi could provide a relatively easy way of measuring how well a group is working together and using that to predict how well the group will perform on other tasks in the future.

Of course, it is certainly possible that other metrics would have predictive power similar to that of phi. Therefore, we believe an important task for future research is to investigate the predictive power of various other metrics. For instance, it is possible that some of the information theoretic or correlational quantities used to compute phi would, themselves, predict performance as well as phi does. Or, perhaps, other measures of complexity (e.g., [38–41]) would be better predictors. And it will certainly be important to compare the predictive power of phi (or its components) with other potential explanatory variables such as (a) the relative participation of different group members [31], (b) the amount of effort and ability members devote to the group's tasks [42] and (c) different measures of the network topology of the group's interactions [4].

*Measuring the complexity of group interaction*

In addition to predicting group performance in two cases, phi also provided quantitative confirmation of the plausible hypothesis that the Internet is increasing in a particular kind of complexity over time. This kind of complexity involves both increasing differentiation among parts and also increasing integration among the parts. And, as we saw, it is not simply a result of the increasing size of the Internet, since the comparisons all included the same number of nodes.

*Measuring the consciousness of groups*

We conclude this discussion with some speculations suggested by the interpretation of phi in the way it was originally intended—as a measure of consciousness. Of course, our results don't prove that groups can be conscious. That is primarily a philosophical question, not an empirical one. But our results do suggest that regarding groups as conscious, in the sense measured by integrated information, would lead to conclusions that are at least consistent with some of our intuitions about how conscious entities behave. Humans, for instance, generally perform better on a variety of tasks when they are more conscious, and we found a similar result for groups. One might also speculate that, if there is a sense in which the Internet is conscious, then its level of consciousness should be increasing over time. Our results are consistent with this speculation as well.

From a philosophical perspective, the question of whether groups should be considered conscious involves whether doing so would lead to consequences that are unintuitive or otherwise undesirable. For instance, two recent philosophical papers take opposite positions on this question. List [43] argues that, even though groups may well be conscious in the functional sense of awareness, they do not have the subjective experience of consciousness. He supports this view, in part, by speculating that the integrated information of most groups would probably be much lower than that of individuals. He also argues that if we consider groups as conscious, then we might be morally obligated to give them certain legal rights formerly reserved for individuals (such as the free-speech rights in the US Supreme Court's Citizens United decision).

Schwitzgebel [44], on the other hand, argues that *not* allowing the possibility that groups could be conscious leads to conclusions that are incompatible with other materialist assumptions about the world. For instance, he argues that, unless you believe consciousness involves some kind of immaterial spirit, then it should be possible, in principle at least, for it to exist in animals, in hypothetical alien creatures, and in groups of humans. Believing otherwise, he asserts, leads either to inconsistencies or to arbitrarily excluding groups from being conscious simply by definition.

The developers of integrated information theory have taken a philosophical position on this question that is referenced by both List and Schwitzgebel. Tononi and colleagues [12,16,45]) include, in their most recent version of the theory, an "exclusion postulate," which says that we should only regard a system as "conscious" if it is at the level of aggregation that maximizes phi. For example, if the phi value for a human brain is greater than the phi value for a group in which that human participates, then we should regard the human as conscious but not the group.

Of course, it's possible to define consciousness in this way, and doing so avoids a possibility that many people find unintuitive: that groups might be conscious. But intuition is not an infallible guide here. As Huebner [46] writes: "it is hard to imagine that collectivities can be conscious; but it is just as hard to imagine that a mass of neurons, skin, blood, bones, and chemicals can be ... conscious."

And the exclusion postulate has the undesirable consequence that a system can become unconscious, even without any change in its value of phi, if another system that contains it (or that it contains) attains a slightly higher value of phi. For instance, imagine that humans will someday be part of a very complex participatory democracy in which people are massively connected online. According to the exclusion postulate, if the participatory democracy becomes more and more richly integrated and someday reaches a level of phi greater than that of the individual humans, then all the individual humans would suddenly become "unconscious." [16,44].

From a scientific point of view, one interesting way to proceed might be to formulate an alternative version of integrated information theory that substantially weakens or eliminates the exclusion postulate. Such a theory might, for instance, allow us to say that conscious humans could also be part of higher-level groups that had a certain level of consciousness themselves. And it seems possible, at least, that such a theory could lead to interesting insights about groups as well as individuals. For example, as List [43] points out, corporations often act as more efficient profit maximizers than individual humans, and perhaps viewing the corporations as doing this consciously would provide useful insights.

**Conclusion**

In this paper, we have seen how the mathematical concept of integrated information formalizes observations about the importance of differentiation and integration that have arisen, more or less independently, in a number of different disciplines. We have also seen how applying this metric to empirically analyze group interactions can lead to potentially useful predictions of group performance, measurements of group complexity, and measurements of something we might even want to call group consciousness.

Much work remains to be done, but, perhaps, applying the concept of integrated information to large groups will be especially useful in understanding the complex kinds of hybrid human-computer systems that are becoming increasingly important in our modern world.

## Methods

The research was approved by Massachusetts Institute of Technology's Institutional Review Board.

**Study 1: Small work groups**

*Measuring the collective intelligence of small work groups*

The data about collective intelligence were collected during a previous study that tested the impact of mode of communication on general group performance [30]. In this study, groups of four people worked together on a set of diverse online tasks. The tasks included both verbal and nonverbal activities of the following types: generating, choosing, remembering, sensing, and taking physical actions. For instance, tasks included brainstorming uses for a brick, solving Raven's Matrices problems from a standardized intelligence test, remembering features of complex videos and images, and copying complex text passages into a shared online editor. Detailed task descriptions and descriptive statistics are included in [30], and summary task descriptions are in S1 and S2 Tables.

All group members used individual laptop computers to work on the shared online tasks. In one condition, the group members were seated near each other and were able to communicate face-to-face while solving the tasks. In the other condition, the group members were seated far apart and were only able to communicate via the text chat functionality built into the online system.

To determine the collective intelligence scores for the different groups, we performed a factor analysis of the groups' scores on the different tasks. As with previous work [31], the first factor in these analyses explained around 40% of the variance in the groups' performance on all the tasks. We treated each group's score on this first factor as the group's collective intelligence. This collective intelligence score, therefore, is a weighted average of the group's scores on all the tasks with the weights chosen to maximize the predictive power for performance on all the tasks. In this sense, the collective intelligence score for a group is exactly analogous to individual intelligence test scores for individuals [31,37].

*Calculating phi for small work groups*

For analyzing the data from Study 1, we used the phi metric that Barrett and Seth [28] call $\Phi_E$ ("empirical phi"). This metric is based upon the theoretical definition of phi by Balduzzi and Tononi [13] and assumes that the system being analyzed is stationary. It can be written as:

$$\Phi_E[X; \tau] = I(X_{t-\tau}; X_t) - \sum_{k=1}^{2} I(M^k_{t-\tau}; M^k_t)$$

where *X* is a stochastic system, $\tau$ is the time delay with respect to which phi is measured, $X_t$ is the state of the system at time *t*, and $M^1$ and $M^2$ are subsets of *X* chosen such that they constitute a minimum information bipartition (MIB) of *X*.

*I* (*X,Y*) is the mutual information between *X* and *Y* which is defined as the reduction in uncertainty (entropy), about X, knowing the outcome of Y:

$$I(X; Y) = H(X) - H(X|Y).$$

Thus $\Phi_E$ is another way of calculating the information generated by the system as a whole that is more than just the sum of its parts.

To use this metric, we recorded communication in different ways for the two conditions. For the face-to-face condition, each group member had an individual microphone. This resulted in four time-aligned audio tracks. We first used software to split the audio tracks into time steps of 200 ms each. The software then determined for each time step, who, if anyone, was speaking. To do this, the software analyzed which group members' audio volumes were above a threshold level. This level was optimized based on "ground truth" data obtained from human observer ratings of who was speaking for a limited subset of the data. The next step suppressed the audio tracks that picked up muted versions of someone else's speech. The final step merged speaking turns of a single speaker that were 400ms or less apart (e.g. someone making a brief pause during a speaking turn).

This procedure thus yielded, for each team, a state vector that encoded everyone who was speaking at a specific time step with a 1 and everyone else with a 0. We then applied the phi metric to this state vector.

For the online condition, we used software to analyze the chat transcripts. We encoded each line of chat as one time step. During this time step, the group member who chatted is encoded as 1 (for active) and all other group members are encoded as 0 (for inactive). This encoding leads to a situation (unlike with the face-to-face groups) where only one person can be active at any given time. We then computed the phi metric on this state dataset.

Note that, in this study and the other two, we assume, but do not test for, stationarity of the time series of state vectors that are used to calculate phi. We believe that the results reported above about correlations between phi and other variables are of interest, in any case, even if they are caused, in part, by factors that led to non-stationarity in the systems. However, as noted in the Discussion above, we also believe that an important focus for future work would be to examine many alternative factors that might explain our results, including any that might have involved non-stationarity in the systems.

**Study 2: Groups of Wikipedia editors**

*Measuring quality of Wikipedia articles*

In this study, we analyzed the edit histories of the articles in Wikipedia's Vital Articles list [32]. At the time of downloading, this included 1000 articles spanning a wide range of rated quality levels, topics, and popularity. We discarded the Wikipedia front-page article since it had an order of magnitude more edits than any other article in the list and thus was as a clear outlier. This left 999 articles that we analyzed.

From the edit histories of these articles, we parsed the quality level of the articles for each edit step giving us the points in time when changes in quality occurred. We then analyzed the 30-, 60-, and 90-day periods before each quality change, discarding the periods in which only one or two editors were active (see descriptive statistics in S3 Table).

*Calculating phi for groups of Wikipedia editors*

We computed phi for each article in the 30-, 60-, and 90-day periods before each quality change. As with the chat transcripts in Study 1, we encoded each edit as a single time step. An editor was considered to be active if he or she edited the Wikipedia article in question at that time step and inactive otherwise.

However, we could not compute phi for Study 2 using the $\Phi_E$ metric used in Study 1 for two reasons. First, as the number of nodes in the network grows, it becomes increasingly difficult to obtain enough data to accurately estimate all the relevant entropies using $\Phi_E$ [28]. To deal with this problem, we used the phi metric that Barrett and Seth call $\Phi_{AR}$ ("auto-regressive phi"). This metric provides reasonable estimates for both Gaussian and non-Gaussian systems with smaller amounts of data [28] and can be written as:

$$\varphi_{AR}[X; \tau, \{M^1, M^2\}] = \frac{1}{2}\log\left\{\frac{\det \Sigma(X)}{\det \Sigma(E^X)}\right\} - \sum_{k=1}^{2} \frac{1}{2}\log\left\{\frac{\det \Sigma(M^k)}{\det \Sigma(E^{M^k})}\right\}$$

where $M^1$ and $M^2$ are a bipartition of the data, $\det\Sigma(X)$ is the determinant of the covariance matrix of X, and $E^{M^k}$ and $E^X$ are residuals in regression equations that estimate states of the system at one time based on knowledge of the system state at another time. To compute this version of phi, we used a MATLAB toolbox provided by Adam Barrett [28].

The second problem that arises with large systems is that determining the minimum information bipartition (MIB) requires enumerating all possible bipartitions of the dataset. Since the number of these bipartitions grows exponentially with the size of the network, this method quickly becomes computationally infeasible. To avoid these problems, we used "atomic" partitions in determining phi as recommended by [29,47]. With this approach, each node is considered as its own partition $M^k$ and the summation in the second term is done over all these.

We verified the validity of this atomic measure on our dataset by computing the normal phi and the atomic phi for all edit histories from the Wikipedia dataset with 14 editors or less, 14 being the largest number where enumerating all bipartitions is still computationally feasible for all articles. The values of the atomic phi are higher but the correlation between the original phi and the atomic phi was highly significant (r = 0.83, p<0.001).

In some cases in our data, the $\Phi_{AR}$ algorithm became numerically unstable and was unable to return a value at all or returned a value that was theoretically impossible (that is, less than 0 or greater than the number of nodes). These problems usually occurred in cases where many nodes had (little or) no variance in their activities (e.g., the nodes were almost always on or almost always off). In these cases, the state vector matrices often became rank deficient, and the algorithm was unable to compute a valid value for phi.

Since nodes with (little or) no variance have (little or) no entropy, they also have (little or) no effect on phi. Therefore, in cases where the algorithm did not return a valid value for phi, we simply dropped the 5% of nodes with the least variance and reran the computation, repeating this procedure until the algorithm did return a valid value. In Study 2, these problems occurred in 19.5% of the cases, and when they occurred, we had to repeat the procedure 2.27 times on average. In other words, we corrected for numerical instabilities in the calculation of phi by removing a small number of low-variance nodes that would have had little effect on phi in any case. This insured that all the data analyzed was for groups of nodes for which valid values of phi could be computed.

**Study 3: Groups of computers and people on the Internet**

To analyze Internet traffic, we used a database compiled by the Cooperative Association for Internet Data Analysis (CAIDA) [34]. This database includes records of the data logged by two high-speed monitors on a commercial backbone link on the Internet. The monitors are in Chicago and San Jose, and we chose the one in San Jose since it provides a longer undisrupted history (from 2008 to the present). We analyzed datasets separated by approximately 6-month intervals during the period (usually every March and September).

Since the volume of Internet backbone traffic is huge, the database includes only one hour of data for each month, and we limited our analysis to one minute of this data for the months we analyzed. We picked the fourth minute of each hour to avoid any unusual activities in the first minute of the hour (such as special programs that operate automatically at the beginning of each hour).

The database contains a trace of each packet of information sent, including an anonymized version of the Internet Protocol (IP) address for the origin and destination node of each packet. Each node (or "host") is a different computer, such as an end user's laptop, a mail server, or a web server for Google, Amazon, and other web service providers. The IP addresses for these nodes are anonymized in such a way that each real IP address always matches to the same anonymized counterpart.

Descriptive statistics for the dataset, after unpacking and parsing (including, for example, removing IPv6 and unreadable packets) are shown in S4 Table.

*Calculating phi*

We calculated phi for Study 3 using the same phi metric used for Study 2. To do this, we characterized the state of the system in terms of which nodes were active (in the sense of sending an information packet) at a given time. We also determined a time delay with respect to which to calculate phi. In order to do these things, several other steps were also needed.

*Sampling nodes*

As shown in S4 Table, the number of nodes sending packets in the months we analyzed ranged from about 200,000 to 1.6 million. We know of no method for calculating phi that is computationally feasible and numerically stable for systems with anything remotely approaching this number of nodes, so, before calculating phi, we needed to subsample the nodes to be analyzed. Ideally, these sampling methods should select subsets of nodes whose activity relationships are representative of those in the whole sample. Therefore, the first two methods we used were the two methods for sampling from large graphs that were found by Leskovec and Faloutsos [48] to best retain the network properties of the graphs.

In describing these methods, we denote by **S** the set of all nodes in a sample of information packets, and by **A** ⊂ **S** the subsample of nodes to be analyzed. Bold lower case letters indicate single nodes. D(**a**) refers to the set of all destination nodes to which node **a** sent a packet in the sampled period and D(**A**) is the set of all destinations for any node in **A**. We do not allow duplicate nodes in **A**.

The two methods we used were:

a) *Random walk*. We first pick a random node **x** ∈ **S**, add it to **A**, and make it the active node. We then randomly pick a new active node **y** ∈ D(**x**) and add it to **A**. At each step, we continue by doing one of two things. With probability 0.85, we pick a new active node from the destinations of the current active node. And with probability 0.15, we return to **x** and start a new path from there. If we run out of new nodes to visit (e.g. in the case of a small isolated subset) we pick a new starting node **x**. This method is repeated until we have reached the sampling goal. As stated by [48] the return probability of 0.15 is the standard value picked in literature.

b) *Forest fire*. We first pick a random starting node **x** ∈ **S**, add it to **A**, and make it the active node. Next we pick a random number *n* from a geometric distribution with mean 2.3 (the value suggested by [48]), randomly pick *n* nodes from D(**x**), and add these nodes to **A**. The procedure continues by selecting new active nodes from **A** and repeating the process until the required number of nodes is reached. If at any point, there are no nodes left in D(**A**) that are not already in **A**, then a new random starting node **x** ∈ **S** is selected and added to **A**.

For comparison, we also used two other simple sampling methods:

c) *Breadth first*. We randomly pick **x** ∈ **S** as our starting node and add it to **A**. We then iteratively add to **A** all nodes to which nodes in **A** sent packets (i.e. D(**A**)) until we reach our sampling goal. If there are no more nodes in D(**A**) that are not already in **A**, we pick a new starting node **x** ∈ **S** and continue from there.

d) *Random nodes*. We randomly pick **x** ∈ **S** and add it to **A** until we reach our sampling goal. Note that this method selects a small number of nodes (e.g., 100) completely randomly from a much larger set (e.g., several hundred thousand nodes). Therefore, even if there are substantial interactions among nodes of the type phi measures, this node sampling method may not detect them very well. However, we still include it for comparison purposes.

For each date and each node sampling method, we created 100 different random subsamples of nodes. We then computed phi on the resulting state vectors and averaged the results across all 100 different random subsamples.

*Determining time step size and time delay*

To characterize the state of the system, we needed to determine the size $\delta$ of the time steps into which activity data will be grouped (i.e. for which we assume all the data packets are sent at the same time). We also needed to pick a time delay $\tau$ with respect to which phi will be calculated. These two factors depend on each other logically. For instance, if there are true interactions at a timescale of 100 ms, we could detect them with phi by, for example, setting $\delta$ = 100 ms and $\tau$ = 1 time step or by setting $\delta$ = 50 ms and $\tau$ = 2 time steps.

To make the search space of possibilities more manageable, we fixed the time delay $\tau$ = 1 and selected the time step size $\delta$ that maximized phi when averaged across all the dates in our analysis. In calculating phi for this purpose, we made the following assumptions: (a) node sampling was done using the random walk method, and (b) the other corrections described below were made. This resulted in a time step size $\delta$ of 100 ms (see Fig 3). We also obtained similar results for other combinations of parameters.

This corresponds very well with typical response times observed on the Internet. As noted by [49], the typical "round trip" time for data on the Internet to travel from point A to point B and back is about 200 ms. If we make the reasonable assumption that the processing time on the remote machine is minimal, then the delay is almost entirely due to time spent travelling back and forth on the network, so each one-way trip would be about 100 ms. The time delay relevant for calculating phi is the delay for one-way travel plus the time for the remote machine to respond, so these numbers correspond very well.

*Determining node sample size*

Based on preliminary experiments with our data, we found that the computations for phi often became numerically unstable and very computationally expensive at around 200 nodes. To avoid these problems we picked a standard node sample size of 100 nodes. As noted below, however, the results were also similar with samples of 150 and 200 nodes.

*Correcting for numerical instabilities*

We used the same method to correct for numerical instabilities as used in Study 2. In Study 3, invalid values occurred initially in 67.14% of the cases, but they disappeared after repeating an average of 2.12 times the procedure of dropping low variance nodes.

*Correcting for hardware change at the recording site*

As mentioned in the main text, the hardware at the recording site was upgraded in the time period between September 2011 and March 2012, which led to a noticeable drop in the phi values. To correct for this, we added an indicator variable to our linear model that indicates if the date is before or after March 2012. This allowed us to extrapolate the corrected phi according to the model. For readability reasons, the graphs in Fig 4 and S2-S4 Figs show the extrapolated values in red, the uncorrected values in light red, the regression line for the corrected values as a red line and the regression line for the uncorrected values as a black line.

*Results*

Using the procedures just described, we calculated the value of phi over time for four node sampling methods (random walk, forest fire, breadth first, and random nodes). The resulting graphs are shown in Fig 4 and S2 Fig. In all cases except random node sampling, the relation between phi and year is positive and very significant (see Table 2). As noted above, we did not expect the random node sampling method to be very effective at detecting interactions of the sort phi measures, so it is not surprising that the results were not significant in this case.

|  | Node sampling method | | | |
| --- | --- | --- | --- | --- |
|  | **Random Walk** | **Forest Fire** | **Breadth First** | **Random Nodes** |
| Regression coefficient | 1.675*** | 1.676*** | 1.715*** | -0.26 |

\*\*\* = p < $10^{-8}$

**Table 2.** Regression coefficients for predicting phi from date with four different node sampling methods

*Robustness check for time step size*

As noted above, the main results were calculated with a time step size $\delta$ = 100 ms which maximized the value of phi. However, S3 Fig shows that using time step sizes of 50 ms or 150 ms also yields similar results.

*Robustness check for node sample size*

As noted above, the main results were calculated with a node sample size of 100 nodes. However, S4 Fig shows that using sample sizes of 150 or 200 also yield similar results.

*Robustness check for number of packets sampled*

As shown in S4 Table, the number of packets sent in the minutes we studied is not constant over the dates we studied. To investigate whether the variable number of packets could have affected the results, we also investigated a different method for sampling packets. With this alternate method, we analyzed only the first 10,000,000 packets in each minute, since this is the maximum (round) number of packets present for all dates. S5 Table shows the regression coefficients for this sampling method. We see again that date is a significant predictor of phi in this case for all four sampling methods.

**Acknowledgments**

This work was made possible by financial support from the National Science Foundation (grant numbers IIS-0963285, ACI-1322254, and IIS-0963451), the U. S. Army Research Office (grant numbers 56692-MA, 64079-NS, and W911NF-15-1-0577) and Cisco Systems, Inc., through their sponsorship of the MIT Center for Collective Intelligence. We wish to especially thank Adam Barrett from the University of Sussex for providing the MATLAB toolbox to compute phi. We also thank Stephan Gade for help with the statistical analysis and Larissa Albantakis, Adam Barrett, Tomaso Poggio, and two anonymous referees for providing feedback on previous versions of the paper.

# SUPPORTING INFORMATION

**S1 Table.** Task categories and verbal vs. non-verbal dimensions in the Collective Intelligence task battery (reproduced from (23))

| Task Category | Verbal | Non-Verbal |
| --- | --- | --- |
| 1. Generating | Brainstorming Words | Brainstorming Uses for a Brick<br>Brainstorming Equations |
| 2. Choosing | Unscramble Words<br>Judgment Slogans | Matrix Reasoning<br>Sudoku<br>Judgment Picture<br>Judgment Pages |
| 3. Executing | Typing Text | Typing Numbers |
| 4. Remembering | Memory Words* | Memory Video<br>Memory Images |
| 5. Sensing | Detection Words | Detection Images |

* Due to technical problems with administration of the "Memory Words" task, it was excluded from the analysis.

**S2 Table.** Description of tasks used to measure collective intelligence of groups.

| Task Type | Description |
| --- | --- |
| Brainstorming | Groups had to collectively come up with as many as possible of the following: (a) uses for a brick, (b) words that started with S and ended with N, and (c) equations that equal 10 with certain constraints on operators and values used. Points were given for number of answers and uniqueness of answers. |
| Unscrambling | The subjects were awarded points for correctly identifying words whose letters were randomly scrambled |
| Matrix Reasoning | Solving a set of Raven's Advanced Progressive Matrices, a standardized test of general fluid intelligence. |
| Sudoku | Solving a Sudoku puzzle. Points were awarded for number of correct answers |
| Judgment | Groups had to predict how a larger population would rate the quality of images and slogans. They also had to estimate the number of pages in a book based on a picture of the book. |
| Typing | The groups had to copy a complex text passage and a complex series of numbers into a shared workspace similar to Google Docs. Scoring was based on number of items copied correctly with significant penalties for incorrect and skipped items. Therefore, it was important for groups to carefully coordinate their work to avoid duplications and long sequences of skipped items. |
| Memory | Groups were shown complex videos, images, and sequences of words and then asked to answer a set of questions about the items they had seen. |
| Detection | Groups answered questions about a grid of small images such as "What is the most frequent object in the grid?" |

**S3 Table**: Descriptive statistics of the number of articles, editors, edits, and edits per editor in various periods before quality changes in the Wikipedia dataset (time windows of 30-, 60-, and 90-days shown in panels A, B, and C, respectively).

|  |  | FA | A | GA | B | C |
|---|---|---|---|---|---|---|
| Number of articles |  | 136 | 108 | 185 | 535 | 169 |
| Number of editors per article | Mean | 16.206 | 12.806 | 10.984 | 9.164 | 6.728 |
|  | Min | 2 | 3 | 3 | 2 | 2 |
|  | Max | 72 | 56 | 63 | 51 | 18 |
|  | Std | 12.075 | 9.216 | 8.589 | 7.111 | 3.382 |
| Number of edits per article | Mean | 201.544 | 94.241 | 110.676 | 50.206 | 34.817 |
|  | Min | 4 | 8 | 8 | 4 | 6 |
|  | Max | 1049 | 861 | 825 | 857 | 318 |
|  | Std | 210.452 | 136.702 | 137.086 | 78.652 | 44.514 |
| Number of edits per editor per article | Mean | 12.153 | 8.039 | 10.909 | 5.356 | 5.611 |
|  | Min | 1.75 | 1.833 | 1.8 | 1.667 | 1.75 |
|  | Max | 72.8 | 172.2 | 71.8 | 159.2 | 58.333 |
|  | Std | 11.186 | 17.157 | 11.857 | 9.422 | 8.765 |

(A) 30-day period before quality change

|  |  | FA | A | GA | B | C |
|---|---|---|---|---|---|---|
| Number of articles |  | 138 | 129 | 215 | 699 | 273 |
| Number of editors per article | Mean | 25.92 | 20.163 | 16.893 | 13.701 | 9.183 |
|  | Min | 3 | 3 | 2 | 2 | 2 |
|  | Max | 115 | 89 | 97 | 77 | 31 |
|  | Std | 19.411 | 16.355 | 13.851 | 11.39 | 5.611 |
| Number of edits per article | Mean | 326.768 | 152.24 | 170.995 | 79.495 | 45.388 |
|  | Min | 9 | 7 | 8 | 4 | 4 |
|  | Max | 2341 | 1637 | 969 | 1635 | 350 |
|  | Std | 345.283 | 213.7 | 175.6 | 124.957 | 52.16 |
| Number of edits per Editor per article | Mean | 12.655 | 9.036 | 12.052 | 5.66 | 4.931 |
|  | Min | 2 | 1.75 | 2 | 1.75 | 1.75 |
|  | Max | 70.939 | 272.833 | 104 | 272.5 | 47.5 |
|  | Std | 10.908 | 24.181 | 13.133 | 11.846 | 6.19 |

(B) 60-day period before quality change

|  |  | FA | A | GA | B | C |
|---|---|---|---|---|---|---|
| Number of articles |  | 139 | 135 | 226 | 773 | 345 |
| Number of editors per article | Mean | 34.576 | 26.837 | 22.845 | 18.107 | 11.533 |
|  | Min | 6 | 4 | 3 | 2 | 2 |
|  | Max | 152 | 110 | 121 | 123 | 41 |
|  | Std | 26.284 | 20.8 | 18.114 | 15.609 | 8.222 |
| Number of edits per article | Mean | 413.281 | 203.541 | 229.615 | 106.585 | 54.939 |
|  | Min | 21 | 11 | 9 | 3 | 3 |
|  | Max | 2550 | 1653 | 1721 | 1661 | 420 |
|  | Std | 407.634 | 243.445 | 231.389 | 151.653 | 61.581 |
| Number of edits per editor per article | Mean | 12.466 | 9.051 | 12.47 | 5.647 | 4.591 |
|  | Min | 2.231 | 2.2 | 1.8 | 1.5 | 1.5 |
|  | Max | 67.105 | 236.143 | 245.857 | 237.286 | 46.667 |
|  | Std | 10.067 | 20.515 | 18.921 | 9.684 | 5.049 |

(C) 90-day period before quality change

**S4 Table**: Number of information packets, origin nodes and destination nodes for each month analyzed.

| Year | Month | Number of information packets | Number of origin nodes | Number of destination nodes |
|------|-------|-------------------------------|------------------------|------------------------------|
| **2008** | 8 | 10,419,879 | 284,987 | 248,892 |
| **2009** | 3 | 15,640,702 | 373,440 | 551,702 |
| **2009** | 9 | 17,402,179 | 495,864 | 622,101 |
| **2010** | 3 | 15,322,907 | 343,577 | 399,145 |
| **2010** | 9 | 14,570,116 | 294,594 | 173,678 |
| **2011** | 3 | 18,094,613 | 559,263 | 352,122 |
| **2011** | 9 | 20,132,417 | 411,129 | 412,778 |
| **2012** | 3 | 11,242,632 | 294,890 | 164,138 |
| **2012** | 9 | 10,503,070 | 220,584 | 96,248 |
| **2013** | 3 | 11,195,621 | 1,675,419 | 150,134 |
| **2013** | 9 | 20,739,677 | 274,875 | 204,683 |
| **2014** | 3 | 18,803,582 | 413,410 | 164,051 |

** = $p < 10^{-6}$

**S5 Table**: Regression coefficients for predicting phi from date with four different node sampling methods while only looking at a fixed number of packets (10,000,000 packets).

| | Node sampling method | | | |
|---|---|---|---|---|
| | **Random Walk** | **Forest Fire** | **Breadth First** | **Random Nodes** |
| Regression coefficient | 3.487*** | 3.025*** | 3.850*** | 7.334*** |

*** = $p < 10^{-6}$

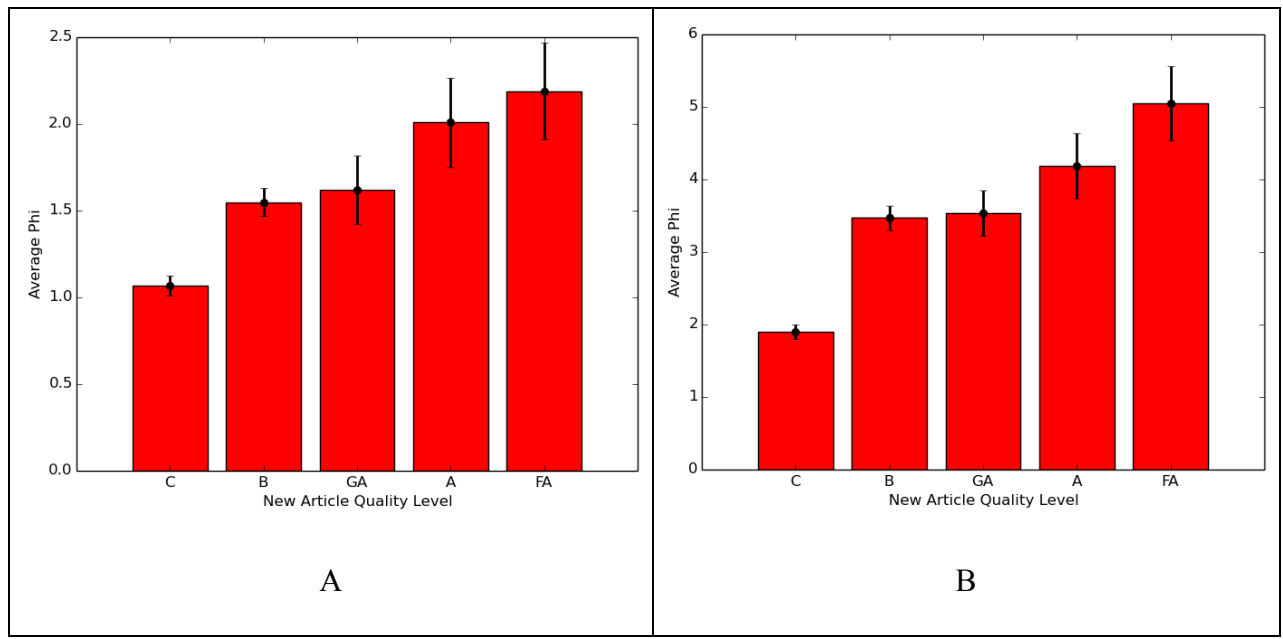

**S1 Fig.** Average phi for groups editing Wikipedia articles of different quality levels in the 30-day period (A) and the 90-day period (B) before the articles were promoted to their current quality level.

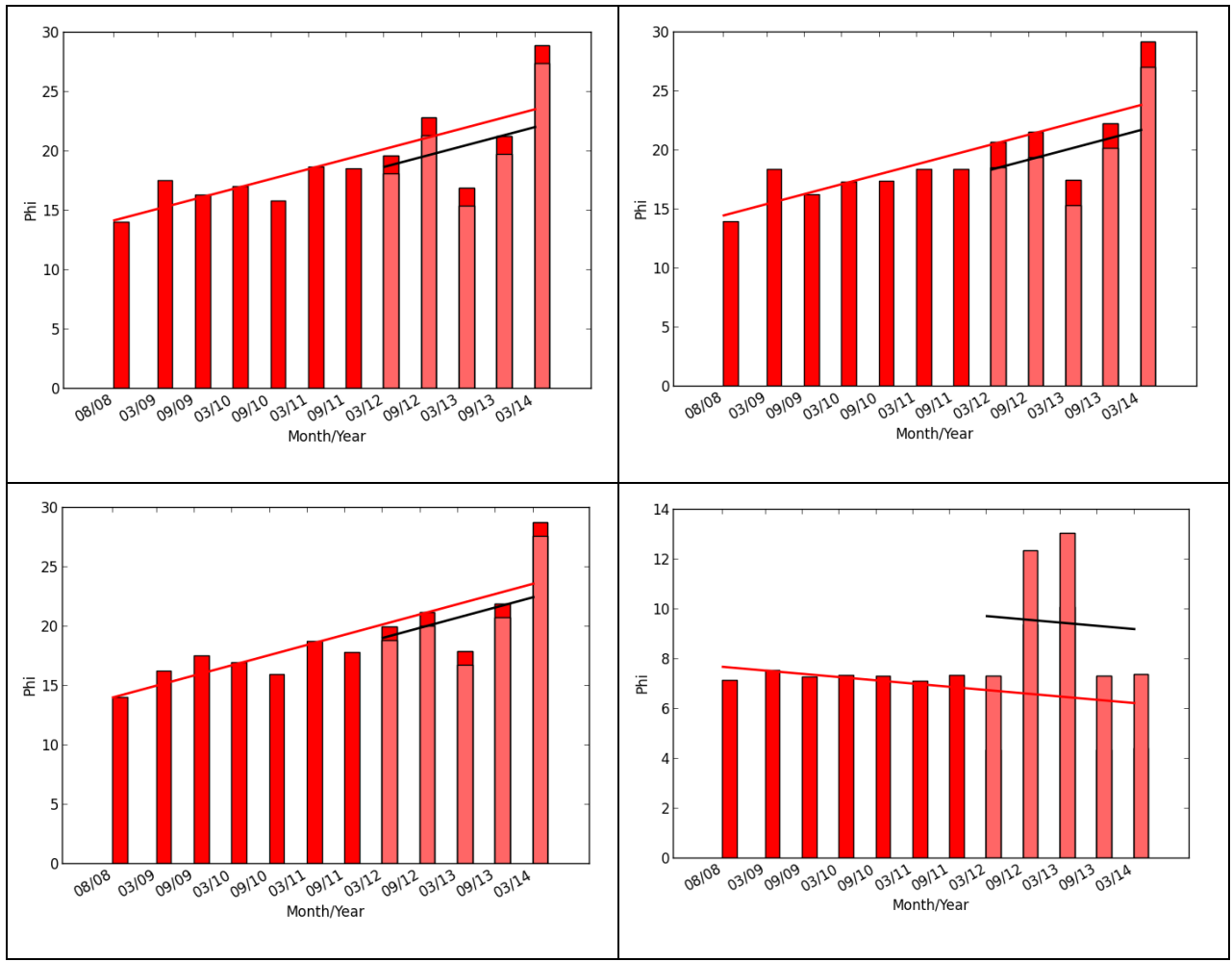

**S2 Fig**: Average phi over time for various node sampling methods (top left to bottom right: Random Walk, Forest Fire, Breadth First and Random Nodes). In all cases node sample size = 100, and time step size $\delta$ = 100 ms.

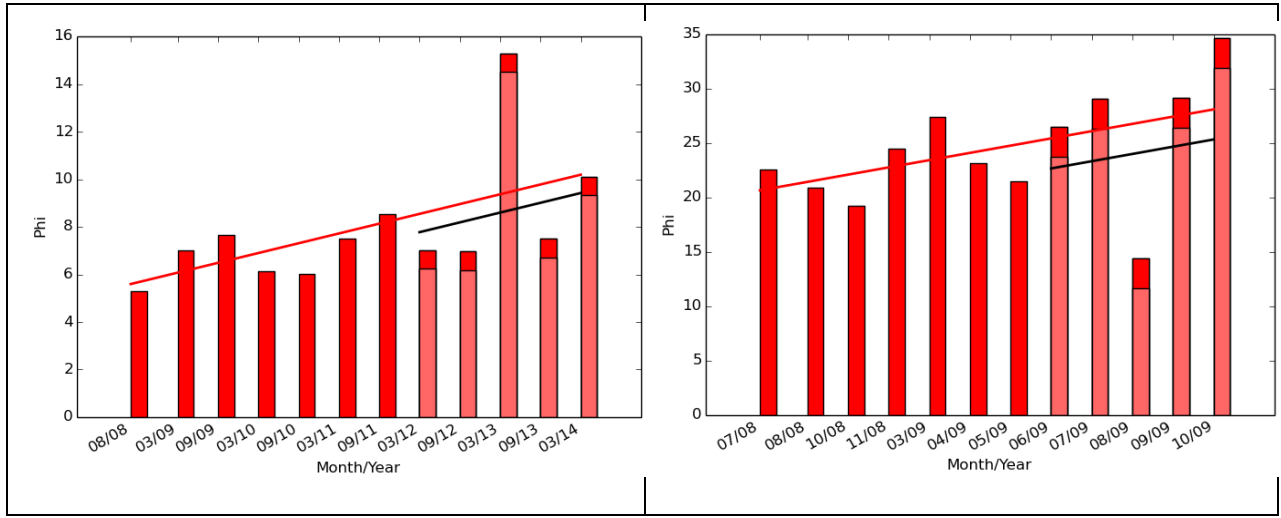

**S3 Fig**: Average phi plotted over time (node sampling = random walk, node sample size = 100, time step size $\delta$ = 50 ms (left) and 150ms (right). For 50ms, $\beta$ = 0.8227, p = 0.000007; for 150ms, $\beta$ = 1.333, p=0.002.

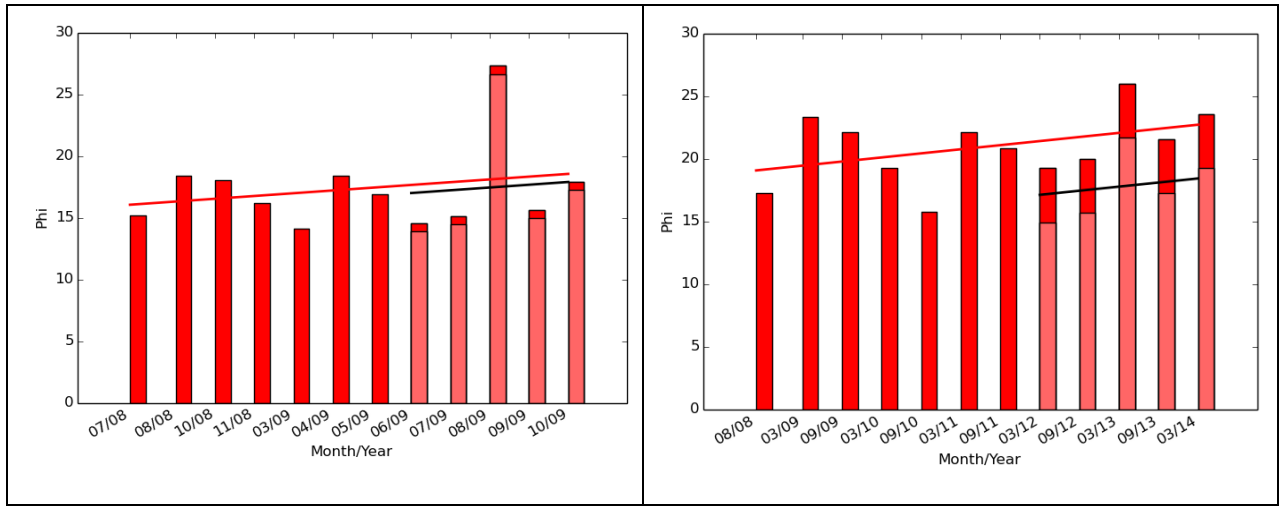

**S4 Fig**: Average phi plotted over time (node sampling = random walk, node sample size = 150 (left) and 200 (right), time step size $\delta$ = 100 ms). For 150 nodes, $\beta$ = 0.4477, p = 0.018; for 200 nodes, $\beta$ = 0.6535, p=0.00026.